\documentclass[aps,prl,
 amsmath,amssymb,
 twocolumn, groupedaddress,
 showpacs]{revtex4-1}

\usepackage{bm}
\usepackage{graphicx}
\usepackage{amsmath}
\usepackage{amssymb}
\usepackage{ulem}
\usepackage{url}

\begin{document}

%\preprint{APS/123-QED}

\title{Fermiology of a topological line-nodal compound CaSb$_2$ and its implication to superconductivity: angle-resolved photoemission study}% Force line breaks with \\

\author{Chien-Wen Chuang,$^{1}$ Seigo Souma,$^{2,3}$ Ayumi Moriya,$^{1}$ Kosuke Nakayama,$^{1,4}$ Atsutoshi Ikeda,$^{5}$ Mayo Kawaguchi,$^{6}$ Keito Obata,$^{6}$ Shanta Ranjan Saha,$^{5}$ Hidemitsu Takahashi,$^{6}$ Shunsaku Kitagawa,$^{6}$ Kenji Ishida,$^{6}$ Kiyohisa Tanaka,$^{7,8}$ Miho Kitamura,$^{9}$ Koji Horiba,$^{9,10}$ Hiroshi Kumigashira,$^{11}$ Takashi Takahashi,$^{1}$ Shingo Yonezawa,$^{6}$ Johnpierre Paglione,$^{5}$ Yoshiteru Maeno,$^{6}$ and Takafumi Sato$^{1,2,3,12}$}
%\email[$^{\ast}$Corresponding author: ]{t-sato@arpes.phys.tohoku.ac.jp}
\affiliation{$^1$Department of Physics, Graduate School of Science, Tohoku University, Sendai 980-8578, Japan\\
$^2$Center for Science and Innovation in Spintronics (CSIS), Tohoku University, Sendai 980-8577, Japan\\
$^3$Advanced Institute for Materials Research (WPI-AIMR), Tohoku University, Sendai 980-8577, Japan\\
$^4$Precursory Research for Embryonic Science and Technology (PRESTO), Japan Science and Technology Agency (JST), Tokyo, 102-0076, Japan\\
$^5$Maryland Quantum Materials Center and Department of Physics, University of Maryland, College Park, Maryland 20742-4111, USA\\
$^6$Department of Physics, Kyoto University, Kyoto 606-8502, Japan\\
$^7$UVSOR Synchrotron Facility, Institute for Molecular Science, Okazaki 444-8585, Japan\\
$^8$School of Physical Sciences, The Graduate University for Advanced Studies (SOKENDAI), Okazaki 444-8585, Japan\\
$^9$Institute of Materials Structure Science, High Energy Accelerator Research Organization (KEK), Tsukuba, Ibaraki 305-0801, Japan\\
$^{10}$National Institutes for Quantum Science and Technology (QST), Sendai 980-8579, Japan\\
$^{11}$Institute of Multidisciplinary Research for Advanced Materials (IMRAM), Tohoku University, Sendai 980-8577, Japan\\
$^{12}$International Center for Synchrotron Radiation Innovation Smart (SRIS), Tohoku University, Sendai 980-8577, Japan\\
}
\date{\today}

\begin{abstract}

We performed angle-resolved photoemission spectroscopy with micro-focused beam on a topological line-nodal compound CaSb$_2$ which undergoes a superconducting transition at the onset $T_{\rm{c}} \sim$ 1.8 K, to clarify the Fermi-surface topology relevant to the occurrence of superconductivity. We found that a three-dimensional hole pocket at the $\Gamma$ point is commonly seen for two types of single-crystalline samples fabricated by different growth conditions. On the other hand, the carrier-doping level estimated from the position of the chemical potential was found to be sensitive to the sample fabrication condition. The cylindrical electron pocket at the Y(C) point predicted by the calculations is absent in one of the two samples, despite the fact that both samples commonly show superconductivity with similar $T_{\rm{c}}$'s. This suggests a key role of the three-dimensional hole pocket to the occurrence of superconductivity, and further points to an intriguing possibility to control the topological nature of superconductivity by carrier tuning in CaSb$_2$.
\end{abstract}

\maketitle
\section{INTRODUCTION} 
Initiated by the discovery of topological insulators (TIs) whose surface or edge hosts gapless states despite the insulating nature of bulk \cite{HasanRMP2010, QiRMP2011, AndoJPSJ2013}, the search for new quantum states of matter characterized by the non-trivial topology is becoming one of major challenges in condensed-matter physics. While TI and topological superconductor (TSC) - a superconducting analog of TI - commonly contain a finite energy gap in the bulk (band gap and superconducting gap, respectively), the classification of band topology has been successfully extended to the systems with gapless bulk excitations, leading to the discovery of topological semimetals (TSMs). The first concrete example of TSMs is the three-dimensional (3D) topological Dirac semimetal such as Na$_3$Bi and Cd$_3$As$_2$ \cite{WangPRB2012, WangPRB2013, NeupaneNC2014, BorisenkoPRL2014, LiuScience2014} where a linearly dispersive bulk valence band (VB) and a conduction band (CB) contact each other at a discrete point in the 3D {\bf k} space. Weyl semimetal is also characterized by the point touching of VB and CB, while the bulk band degeneracy is lifted by the breaking of time-reversal or space inversion symmetry \cite{XuScience2015, LvPRX2015, YangNP2015, SoumaPRB2016, ArmitageRMP2018}. There is another type of TSM called line-nodal semimetal where the crossing point extends one-dimensionally in the {\bf k} space (nodal line or loop), being protected by the crystal symmetry such as mirror reflection or nonsymmorphic glide mirror symmetry \cite{BurkovPRB2011, WuNP2016, BianNC2016, BianPRB2016, YamakageJPSJ2016, GuanSciAdv2016, WengPRB2016, TakaneNPJQM2018, XuPRB2015, SchoopNC2016, TakanePRB2016, NeupanePRB2016, louPRb2016, FangCPB2016, ChenPRB2017}. While such TSMs, or more widely, materials containing Dirac/Weyl or line nodes in the bulk-band structure, are already a useful platform to realize exotic quantum phenomena such as unusual magneto-transport properties, the coupling of nodal electrons to superconductivity is another intriguing target of intensive investigations because such a coupling would provoke the topological superconductivity \cite{KobayashiPRL2015}. However, this essential issue remains highly unexplored, because superconducting materials whose low-energy excitations are characterized by Dirac/Weyl or line-nodal electrons are generally rare \cite{BianNC2016, ZhangScience2018, OrtizPRM2019, AggarwalNM2016}.
    
\begin{figure}[htbp]
\includegraphics[width=86mm]{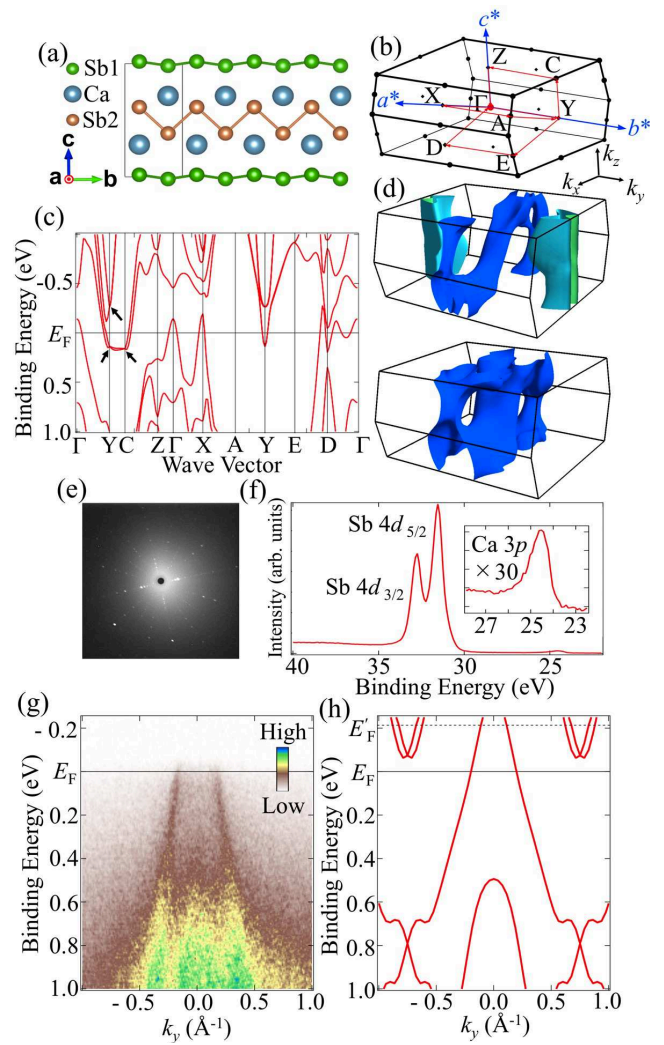}
 \caption{
 \label{Fig1}
(a) Crystal structure (side view) of CaSb$_2$. Two types of Sb atoms at different sites (Sb1 and Sb2) are indicated by green and orange circles. (b) Monoclinic bulk Brillouin zone (BZ) of CaSb$_2$. Red lines indicate high-symmetry lines where the band calculation shown in (c) was carried out. $k_x$, $k_y$, and $k_z$ axes are also indicated. (c) Calculated band dispersions along high-symmetry lines in bulk BZ. (d) Calculated FSs in bulk BZ which incorporate (top panel) no $\mu$ shift and (bottom panel) downward $\mu$ shift of 210 meV. (e) X-ray Laue diffraction pattern of CaSb$_2$ single crystal. (f) EDC which includes the Sb 4$d$ and Ca 3$p$ core levels measured with $h\nu$ = 90 eV. The EDC was obtained by integrating the ARPES spectra collected within angular window of  $\pm$ 15$^{\circ}$ around the surface normal. Inset shows an expanded view of the Ca 3$p$ core level. (g) Near-$E_{\rm{F}}$ ARPES intensity crossing the $\Gamma$ point at $k_z$ = 0 measured at $T$ = 40 K with $h\nu$ = 118 eV. (h) Corresponding calculated band structure along the $\Gamma$Y cut ($k_z$ = 0 plane) which takes into account the downward $\mu$ shift of 210 meV. Dashed line indicates the original Fermi-level position, $E_{\rm{F}}$'.
 		}
 \end{figure}

Recently, the superconductivity with superconducting transition temperature ($T_{\rm{c}}$) of 1.7 K was discovered in a line-nodal compound CaSb$_2$ \cite{IkedaPRM2020}. This material crystalizes in the monoclinic structure with $P2_{1}/m$ space group with Sb zigzag chains running along the $b$-axis and mirror surfaces in the $a$-$c$ plane, as shown in Fig. 1(a), where one can recognize two types of Sb atoms, Sb1 and Sb2, at different sites in the unit cell [note that the bulk Brillouin zone (BZ) is shown in Fig. 1(b)]. First-principles band-structure calculations of CaSb$_2$ predicted the compensated semimetallic band structure [Fig. 1(c)] consisting of a 3D hole pocket extending roughly along the $\Gamma$X direction and quasi-two-dimensional (2D) cylindrical electron pockets axially centered along the YC line [top panel of Fig. 1(d)]. Band structure calculations with the spin-orbit coupling (SOC) predicted the presence of several Dirac points %and line nodes protected by screw symmetry and mirror reflection symmetry, respectively (note that most of them are located in the region above $E_{\rm{F}}$) \cite{FunadaJPSJ2019}. Although the inclusion of SOC leads to the lifting of the degeneracy at these Dirac points, some bands keep the degeneracy 
at specific {\bf k} points such as Y, C, E, and A points (but not $\Gamma$ and Z points which are related to the 3D pocket) due to the nonsymmorphic symmetry \cite{FunadaJPSJ2019} [some of these Dirac points at the Y and C points are indicated by black arrows in Fig. 1(c)]. Such band crossings are connected to the line nodes at the $k_y$ = $\pi$ plane \cite{FunadaJPSJ2019} and one of them is associated with the shallow electron pocket crossing $E_{\rm{F}}$, likely contributing to the transport properties, while these line nodes and Dirac nodes had to await experimental verification.
   
Magneto-transport measurements of CaSb$_2$ clarified a resistivity plateau under a weak magnetic field at 10 K and nonsaturating giant magnetoresistance, consistent with the existence of line nodes \cite{FunadaJPSJ2019}. Electrical resistivity measurements under pressure revealed the dome-shaped $T_{\rm{c}}$ variation without Lifshitz transition in the normal state \cite{KitagawaPRB2021} and specific heat measurements signified the deviation from the BCS behavior \cite{OudahArXiv2021}, suggestive of the unconventional nature of superconductivity. On the other hand, $^{121/123}$Sb nuclear quadrupole resonance (NQR) measurements suggested an exponential decrease in the inverse relaxation rate 1/$T_1$ at low temperature, supportive of the conventional $s$-wave superconductivity with a full gap \cite{TakahashiJPSJ2021}. The presence of cylindrical electron pockets with line nodes has been discussed in favor of topological superconductivity \cite{IkedaPRM2020} associated with the dominant interorbital pairing interaction with odd-parity pairing within the cylindrical FSs, similarly to the case of doped TIs \cite{SasakiPRL2021} and doped Dirac semimetals \cite{KobayashiPRL2015, KawakamiPRX2018}. Topological superconductivity was also predicted based on the full computation of the $Z_2$-enriched symmetry indicators \cite{OnoPRR2021}. Despite such intensive experimental and theoretical studies, there are no experimental reports on the band structure of CaSb$_2$. It is thus of great importance to experimentally clarify the electronic states of CaSb$_2$ to establish the interplay among nodal electrons, superconductivity, and topology.

In this article, we report micro-focused angle-resolved photoemission spectroscopy ($\mu$-ARPES) of CaSb$_2$ to clarify the 3D band structure and Fermi-surface (FS) topology. We observed two different types of FSs in CaSb$_2$; one is a 3D hole pocket at the $\Gamma$ point and another is a cylindrical electron pocket at the Y(C) point.  We found that the carrier-doping level and as a result the FS topology are very sensitive to the sample fabrication condition; the cylindrical electron pocket at the Y(C) point disappears in one of two samples fabricated by different growth conditions, although both samples commonly show superconductivity with similar $T_{\rm{c}}$'s. We discuss implications of the present results in relation to the mechanism of superconductivity and its relationship with topology.         

\section{EXPERIMENTS AND CALCULTIONS}
We prepared two different types of single crystalline samples for ARPES measurements, called here sample A and sample B. The samples A and B were grown by a self-flux method with molar ratios of Ca:Sb=1:5 and 1:3.1, respectively. For sample A, Ca (Sigma-Aldrich, 99.99\%) and Sb (Alfa Aesar, 99.9999\%) were melted at 1000$^{\circ}$C in an alumina crucible inside a sealed quartz tube, cooled down to 740$^{\circ}$C, and then cooled down to 610$^{\circ}$C at a rate of -1$^{\circ}$C/h. Crystals were filtered in a centrifuge at this temperature. For sample B, Ca (Sigma-Aldrich, 99.99\%) and Sb (Rare Metallic, 99.99\%) were heated in a similar temperature profile but in a tungsten crucible. We found that typical size of the single-crystal domains is different between sample A (more than 1 mm size) and sample B (typically a few tens of $\mu$m). From the magnetic susceptibility measurements, we have confirmed the onset $T_{\rm{c}}$ values of samples A and B to be $\sim$ 1.8 K, as detailed in Fig. S1 in Section 1 of Supplemental Material \cite{SM}.

We performed ARPES measurements at BL28A in KEK-PF \cite{KitamuraRSI2022} and BL5U in UVSOR. We used circularly polarized photons of 30-300 eV at BL28A and linear horizontally polarized photons of 20-200 eV at BL5U. The energy and angular resolution was $\sim$ 20 meV and 0.2$^{\circ}$, respectively. The beam size on sample was 10 $\times$ 12 $\mu$m$^2$ \cite{KitamuraRSI2022} and 30 $\times$ 20 $\mu$m$^2$ for BL 28A and BL5U, respectively. Samples were kept at 40 K during measurements. The Fermi-level ($E_{\rm{F}}$) of samples was referred to that of a gold film deposited on the sample holder. Laue backscattering measurements were performed prior to ARPES measurements to determine the sample geometry. A clear Laue pattern was observed for sample A as shown in Fig. 1(e), confirming the high single crystallinity of sample (note that it was difficult to obtain a clear Laue pattern for sample B due to the mixture of multiple crystal domains. However, we could safely determine the sample orientation for a small area of sample B on which the micro-ARPES measurement was performed, by looking at the periodicity and symmetry of obtained band structure). Since CaSb$_2$ single crystal was very hard to cleave, we tried cleaving several times in an ultrahigh vacuum (UHV), occasionally obtained a small flat area on the crystal with a few tens $\mu$m square and then focused the micro photon beam on it. The cleaving plane is $a$-$b$ plane (perpendicular to the $c$* axis) according to our x-ray diffraction and ARPES data. We observed no signature of aging or contamination of sample surface during ARPES measurements.

We have performed the first-principles band structure calculations by using Quantum ESPRESSO \cite{GiannozziJPCM2009} with generalized gradient approximation \cite{GGA} and have included the SOC. The Perdew-Burke-Ernzerhof (PBE) generalized gradient approximation was adopted for the exchange-correlation functional. % We have adopted the full potential linearized augmented plane wave (PAW) method with Perdew-Burke-Ernzerhof (PBE) generalized gradient approximation and have included the SOC. 
 Lattice constants were referenced from the experimental values \cite{IkedaPRM2020} and a $k$ mesh of 18 $\times$ 19 $\times$ 10 was used in the calculations.

\section{RESULTS AND DISCUSSION}
To characterize the sample surface cleaved in UHV, we at first carried out photoemission measurements in a wide energy range with photons of $h\nu$ = 90 eV. The energy distribution curve (EDC) for sample A is shown Fig. 1(f). We observe two peaks at the binding energy ($E_{\rm{B}}$) of 31.5 and 32.7 eV, which are attributed to the spin-orbit-split Sb 4$d_{5/2}$ and 4$d_{3/2}$ orbital, respectively. Single-peaked structure of each orbital suggests that the chemical potential ($\mu$) is almost the same between the Sb1 and Sb2 orbitals. A weak peak at $E_{\rm{B}}$ = 25 eV is assigned to the Ca 3$p$ orbital, as better visualized in the expanded scale in the inset. No other peaks are observed in this $E_{\rm{B}}$ range, supporting the clean surface.

\begin{figure*}
\includegraphics[width=6.7in]{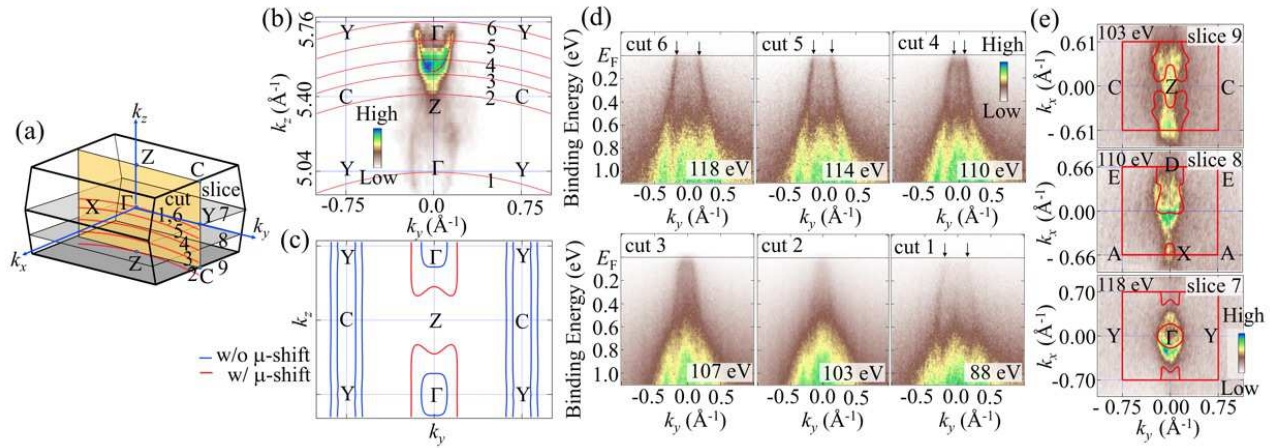}
\begin{center}
 \caption{
 \label{Fig2}
(a) Bulk BZ of CaSb$_2$ with the measured {\bf k} cuts and $k$ planes. Yellow rectangle 
represents the $k$ plane ($k_y$ - $k_z$ plane) where the $h\nu$-dependent out-of-plane FS mapping shown in (b) was performed. Red curves indicate the {\bf k} cuts (cuts 1-6) where the ARPES measurement shown in (d) was carried out. The gray rectangles indicate the {\bf k} slices where in-plane FS mappings shown in (e) were performed. The slices 7, 8, and 9 correspond to $k_z \sim$ 0, -0.5$\pi$, and -$\pi$, respectively. (b) Out-of-plane FS mapping in the $k_y$ - $k_z$ plane obtained by sweeping $h\nu$. Dashed red curves are guide for the eyes to trace the $\Gamma$-centered pocket. (c) Corresponding calculated FSs with (red) and without (blue) taking into account the downward $\mu$ shift of 210 meV. (d) $h\nu$-dependence of experimental band dispersions measured along the {\bf k} cuts (1-6) shown in (a). ARPES data along cuts 1 and 6 nearly trace the same $k$ region in reduced BZ but measured with different $h\nu$'s (118 and 88 eV). (e) In-plane FS mapping obtained with different $h\nu$'s of 118 eV ($k_z \sim$ 0: cut 7), 110 eV ($k_z \sim$ -0.5$\pi$; cut 8), and 103 eV ($k_z \sim$ -$\pi$; cut 9).
		}
 \end{center}
\end{figure*}

Since CaSb$_2$ is a 3D material, it is essential to elucidate the bulk-band structure in the 3D BZ. For this sake, we have carried out $h\nu$-dependent ARPES measurements at the normal emission setup to observe the band dispersion along $k_z$ (parallel to the $c$* axis). We at first estimated the inner potential value to be $V_0$ = 13.7 eV from the periodicity of the band dispersion (for details, see Fig. 2). Then, we performed ARPES measurements at $h\nu$ = 118 eV which corresponds to the $k_z \sim$ 0 plane, along the $\Gamma$Y ($k_y$) cut at $T$ = 40 K. The $\Gamma$Y cut is expected to be useful to simultaneously observe the holelike band forming a 3D pocket at $\Gamma$ and an electronlike band forming a quasi-2D pocket at Y, as suggested from the band calculation displayed in Figs. 1(c) and 1(d). As shown in Fig. 1(g), one can identify a highly dispersive holelike band at $E_{\rm{B}}$ = 0-1 eV which apparently crosses $E_{\rm{F}}$ at $k_y$ = $\pm$ 0.18 \AA$^{-1}$ to form a hole pocket. According to the band calculation \cite{OudahArXiv2021}, this band originates from the Sb2 orbital [see Fig. 1(a)]. Inside this band, another holelike feature is seen at $E_{\rm{B}}$ = 0.5-1 eV. We found that the energy dispersion of these bands shows a qualitatively good agreement with the calculation along the $\Gamma$Y cut shown by red curves in Fig. 1(h), whereas the calculated bands need to be shifted upward as a whole by 210 meV to obtain a best matching in the position of Fermi wave vectors ($k_{\rm{F}}$'s) with the experiment. Surprisingly, we do not observe any signature of electron pockets at the Y point originating from the Sb1 orbital [see Fig. 1(a)] \cite{OudahArXiv2021}, in sharp contrast to the calculation which does not take into account the chemical-potential ($\mu$) shift [note that the $E_{\rm{F}}$ position without the $\mu$-shift ($E_{\rm{F}}$') is indicated by a dashed line in Fig. 1(h)]. This suggests that sample A is hole-doped probably due to the off-stoichiometry of the sample, leading to the lift-up of the 2D cylindrical electron bands into the unoccupied region. We have carefully surveyed the electron pocket at different in-plane ($k_x$ and $k_y$) and out-of-plane wave vectors ($k_z$) and found no evidence for it in the entire 3D BZ as detailed in Fig. 2. Therefore, the transport property of this sample is expected to be governed by the 3D hole pocket, which is schematically illustrated by the calculated FS plot including the hole-doping effect in the bottom panel of Fig. 1(d).

The absence of the electron pocket is suggested from the FS mapping in the $k_y$ - $k_z$ plane [Fig. 2(b)] obtained by changing $h\nu$'s. Although the band calculation without the $\mu$-shift shown in Fig. 2(c) (blue curves) predicted the existence of fairly vertical (i.e. quasi-2D) FSs elongated along the CY cut ($k_y \sim$ 0.75 \AA$^{-1}$), the corresponding ARPES intensity is totally absent as seen in Fig. 2(b). In the experiment, one can see a strong intensity surrounding the top $\Gamma$ point (at $k_z$ = 5.76 \AA$^{-1}$) which produces an ellipsoidal pocket elongated along the $\Gamma$Z direction (dashed red curve). This pocket is associated with the holelike band as seen from the ARPES intensity along the {\bf k} cut crossing the top $\Gamma$ point (cut 6) measured at $h\nu$ = 118 eV in Fig. 2(d). On decreasing $h\nu$ (from cut 6 to cut 4), the separation between two $k_{\rm{F}}$ points indicated by black arrows gradually becomes narrower due to the downward shift of the holelike band. Eventually the $k_{\rm{F}}$ points disappear along cut 3 ($h\nu$ = 107 eV) or cut 2 ($h\nu$ = 103 eV), reflecting the 3D closed nature of the hole pocket as seen from Fig. 2(b). %Due to a mirror symmetry in the $k_x$ = 0 plane, 
We also observe another hole pocket centered at the bottom $\Gamma$ point (at $k_z$ = 5.04 \AA$^{-1}$) which is equivalent to the hole pocket centered at the top $\Gamma$ point, as also seen from the ARPES intensity crossing the bottom $\Gamma$ point along cut 1 in Fig. 2(d). Intensity of this pocket is weaker probably because of the matrix-element effect of photoelectron intensity. One can see from a direct comparison of Figs. 2(b) and 2(c) that the experimental hole pocket is much larger than that in the calculation without the $\mu$-shift (blue curve at $\Gamma$), but becomes comparable after the $\mu$-shift (red curve). This is again due to the aforementioned hole-doping effect.

Now that the absence of 2D electron pocket is established, next we discuss the shape of 3D hole pocket in more detail. For this sake, we performed the in-plane ($k_x$ - $k_y$) FS mapping at representative $k_z$ slices, i.e. $k_z \sim$ 0, -0.5$\pi$ and -$\pi$, as shown in Fig. 2(e). At the slice of $k_z \sim$ 0 ($h\nu$ = 118 eV), one can see a strong intensity concentrated around the $\Gamma$ point which produces a small pocket slightly elongated along the $k_x$ direction. At the $k_z \sim$ -0.5$\pi$ slice ($h\nu$ = 110 eV), this pocket appears to move toward the positive $k_x$ region, causing an apparent asymmetric intensity profile with respect to the $k_y$ (horizontal) axis. At the $k_z \sim$ -$\pi$ slice ($h\nu$ = 103 eV), the intensity shows an hourglass shape with two maxima at top and bottom, distinct from the case of $k_z \sim$ 0 plane where intensity is concentrated at the BZ center. These results suggest that the 3D hole pocket has a rather complicated ``slider'' shape in the 3D BZ, as also inferred from the calculated FS shown in Fig. 1(d). It is noted that the overall $k_z$-dependent evolution of the in-plane FS mapping is consistent with the calculated FS [red curves in Fig. 2(e)]. In particular, the appearance of a small pocket at the $\Gamma$ point at $k_z$ = 0 plane and the movement of this pocket to the positive $k_x$ region with increasing $k_z$ is well reproduced by the calculation.

\begin{figure}[t]
\includegraphics[width=86mm]{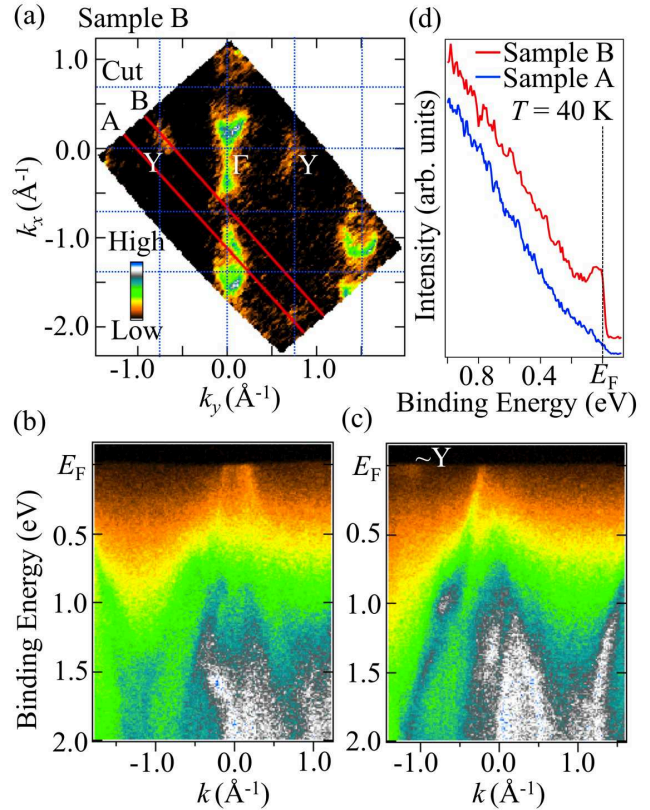}
 \caption{
 \label{Fig3}
(a) FS mapping at $T$ = 40 K measured with $h\nu$ = 90 eV (corresponding to $k_z$ $\sim$ 0 plane) for sample B. Blue lines indicate high-symmetry lines in surface BZ. Red lines indicate representative {\bf k} cuts (cuts A and B) where the ARPES intensity shown in (b) and (c) was obtained. (b), (c) ARPES intensity as a function of wave vector and binding energy measured along cuts A and B, respectively. $k$ in horizontal axis is defined as the distance between the measured $k$ point and the intersection of red line and $k_x$ ($k_y$ = 0) axis. (d) EDC around the Y point which signifies the peak and metallic Fermi-edge cut-off associated with the electron pocket.
}
\end{figure}

To clarify to what extent the observed band structure and fermiology is the intrinsic nature of CaSb$_2$ crystal, we carried out a similar ARPES measurement for sample B. Figure 3(a) shows the in-plane FS mapping at $T$ = 40 K measured at $h\nu$ = 90 eV which corresponds to the $k_z \sim$ 0 plane. One can see a dominant intensity around the $\Gamma$ point elongated along the $k_x$ direction. This feature originates from the Sb2-orbital-derived 3D hole pocket \cite{OudahArXiv2021} as in the case of sample A, but its intensity distribution is different from that of sample A at $k_z \sim$ 0 shown in the bottom panel (slice 7) of Fig. 2(e), likely due to the difference in their doping levels as detailed below. The existence of a hole pocket is recognized from the clear $E_{\rm{F}}$-crossing of holelike band along a representative {\bf k} cut (cut A) shown in Fig. 3(b). Besides the hole pocket, one can see in Fig. 3(a) a weak intensity around the Y point which is well separated from the hole pocket. Intriguingly, this feature is absent in sample A as seen in Figs. 2(b) and 2(e). The ARPES intensity along {\bf k} cut crossing this feature (cut B) in Fig. 3(c) signifies the existence of a shallow electron pocket around the Y point. The metallic electron pocket is also confirmed by the energy distribution curve (EDC) at the Y point shown in Fig. 3(d) in which a peak near $E_{\rm{F}}$ associated with the electron-band bottom accompanied with the Fermi edge cut-off is clearly seen. We have confirmed the existence of electron pocket also at $h\nu$  = 120 eV, in line with its quasi-2D nature predicted from the calculation shown in Figs. 1(c) and 1(d) (for details, see Section 2 of Supplemental Material \cite{SM}). These results indicate that sample B is relatively more electron doped than sample A, and is situated in a semimetallic phase with both hole and electron pockets as predicted by the calculation for stoichiometric CaSb$_2$ (i.e. the calculation without $\mu$-shift). It is thus suggested that the absence of electron pocket in sample A is not due to the experimental artifacts such as the strong intensity suppression associated with the matrix-element effect but due to the difference in the doping levels. This conclusion is also supported by the quantitative analysis of the experimental hole-band dispersion, the photoemission spectra in a wider energy range, and the Sb 4$d$ core-level energies, as detailed in Section 3 of Supplemental Material \cite{SM}.

Now we discuss implications of the present results in relation to superconductivity. Our susceptibility measurements indicate that both samples A and B show superconductivity with almost the same onset $T_{\rm{c}}$ of $\sim$ 1.8 K (see Section 1 of Supplemental Material \cite{SM}), despite the marked difference in the carrier-doping level. The fact that the FS of sample A is composed only of the Sb2-derived 3D hole pocket at the $\Gamma$ point suggests that the main player of superconductivity is the hole carriers in this pocket. It is thus inferred that the electron carriers in the Sb1-derived quasi-2D FS centered at the Y(C) point is not essential for the occurrence of superconductivity. This argument puts a constraint on the microscopic origin of superconductivity as well as its possible non-trivial nature in CaSb$_2$, because the occupancy of the electron band hosting the line nodes was suggested to be important for promoting topological superconductivity associated with the dominant interorbital pairing interaction with odd-parity pairing \cite{IkedaPRM2020}. Also, it is worth noting that CaSb$_2$ maintains superconductivity regardless of the existence or absence of the electron pocket, which points to an intriguing possibility that one can tune the topological nature of superconductivity by simply controlling the carrier concentration in CaSb$_2$, although its validation needs further experimental studies that directly connects the fermiology and the superconducting pairing symmetry. We leave such an experiment as a challenge in future.

\section{CONCLUSION}
  We reported results of a photon-energy-tunable micro-focused ARPES study on a topological line-nodal material CaSb$_2$. We found two different types of FSs; one is a three-dimensional hole pocket at the $\Gamma$ point and another is a cylindrical electron pocket at the Y(C) point which is predicted to host line nodes. When the doping level is changed to a more hole-doped region, the electron pocket at the Y(C) point disappears while the hole pocket at the $\Gamma$ point survives, although the superconductivity keeps to emerge regardless of the doping level. This suggests a dominant role of the 3D hole pocket at the $\Gamma$ point for the occurrence of superconductivity. The present result lays the foundation for understanding the nature of superconductivity and its relationship with topology in CaSb$_2$.

\begin{acknowledgments}
We thank M. Tomikawa for his assistance in the sample characterization. This work was supported by JST-CREST (No: JPMJCR18T1), JST-PRESTO (No: JPMJPR18L7), Grant-in-Aid for Scientific Research (JSPS KAKENHI Grant Numbers JP21H04435, JP20H01847, JP20H00130, JP20KK0061, JP21K18600, and JP22H04933), the US Department of Energy Award for low temperature sample characterization (No. DE-SC0019154), the Gordonand Betty Moore Foundation's EPiQS Initiative for sample synthesis (No. GBMF9071), the Maryland Quantum Materials Center, KEK-PF (Proposal number: 2021S2-001), and UVSOR (Proposal number: 21-658 and 21-679). C.-W. C. acknowledges support from GP-Spin and JSPS.\end{acknowledgments}

%\bibliography{PolarityCVS}% Produces the bibliography via BibTeX.


\end{document}